\documentclass[pdflatex,sn-basic]{sn-jnl}

\usepackage{float}
\usepackage{subfigure}
\usepackage[draft]{minted} 
\usepackage{adjustbox}
\usepackage{makecell}

\usepackage{titlesec}
\titleformat{\subsubsubsection}
  {\normalfont\normalsize\bfseries}{\thesubsubsubsection}{1em}{}
\titlespacing*{\subsubsubsection}
  {0pt}{3.25ex plus 1ex minus .2ex}{1.5ex plus .2ex}

\jyear{2024}%

\theoremstyle{thmstyleone}%
\theoremstyle{thmstyletwo}%

\theoremstyle{thmstylethree}%

\begin{document}

\title[ArcNeural]{ArcNeural: A Multi-Modal Database for the Gen-AI Era}


\author*[1]{\fnm{Wu} \sur{Min}}\email{wumin@hdu.edu.cn}
\author[2]{\fnm{Qiao} \sur{Yuncong}}\email{qiaoyuncong@fabarta.com}
\author[2]{\fnm{Yu} \sur{Tan}}\email{maoqi@fabarta.com}
\author[2]{\fnm{Chenghu} \sur{Yang}}\email{Yang@fabarta.com}

\affil*[1]{\orgdiv{Information Engineering}, \orgname{Hangzhou Dianzi University},  \city{Hangzhou}, \postcode{{310018}},  \country{China}}
\affil[2]{\orgname{Hangzhou Sandeng Tech Inc. \& Fabarta Inc.}, \city{Hangzhou}, \country{China}}


\abstract{ 
The advent of Generative Artificial Intelligence (Gen-AI) and Large Language Models (LLMs) has ushered in a new era of data management, demanding robust and flexible database systems capable of supporting multi-modal data processing. This paper introduces ArcNeural, a novel multi-modal database designed to address the challenges of integrating and managing diverse data types, including graphs, vectors, and documents, in the context of AI-driven applications. Leveraging a storage-compute separated architecture, ArcNeural combines graph technology with advanced vector indexing and transaction processing capabilities to enable efficient data handling and real-time analytics. We present the system’s innovative features, including its unified storage layer, adaptive edge collection in MemEngine, and seamless integration of transaction processing (TP) and analytical processing (AP). Through experimental evaluations, ArcNeural demonstrates superior performance and scalability compared to state-of-the-art systems, making it a versatile solution for enterprise-grade AI applications. This work highlights ArcNeural’s potential to bridge the gap between structured and unstructured data management, paving the way for intelligent, data-driven solutions in the Gen-AI era.}




\maketitle

\section{Introduction}
\subsection{Background and Motivation}
The rapid advancements in Generative Artificial Intelligence (Gen-AI) technology, particularly with the emergence of Large Language Models (LLMs), have marked a significant turning point in the AI era. This new technological paradigm is reshaping the way enterprises develop and deploy AI applications, accelerating their transition towards comprehensive intelligent upgrades. LLMs, with their exceptional generalization capabilities and ability to handle multi-modal AI tasks, are revolutionizing the AI application development and operational maintenance landscape.
However, as enterprises attempt to leverage private data to drive AI applications, they face a series of new challenges. These include ensuring the accurate understanding and utilization of data by AI technologies, maintaining data security, and enabling personalized applications that cater to enterprise-specific requirements. To address these challenges, enterprises need to construct more intelligent and flexible database systems that can effectively support and integrate with large-scale models.
In the era of large models, the effectiveness of AI applications not only depends on the algorithms themselves but also heavily relies on the robustness and flexibility of the supporting data infrastructure and databases. Therefore, designing and implementing a database capable of supporting traditional structured data while efficiently processing and storing various unstructured and semi-structured data becomes crucial. Such a database should be able to comprehensively handle multiple data types, including graphs, text, and vectors, to accommodate the increasingly complex demands of AI applications.
\subsection{Research Objectives}
This paper introduces ArcNeural, a multi-modal database designed to address the challenges and requirements of AI applications in the era of large models. ArcNeural is built on the foundation of graph technology, leveraging its symbolic nature and interpretability to complement probability-based machine learning techniques. By combining the strengths of both approaches, ArcNeural aims to provide a platform for intelligent data understanding and efficient reasoning.
The main objectives of this research are as follows:
\begin{itemize}
\item To present the architecture and key features of ArcNeural, highlighting its ability to efficiently integrate and process multi-modal data, including vectors, documents, and graphs.
\item To discuss the innovative techniques employed in ArcNeural's storage layer, transaction processing (TP) compute nodes, and analytical processing (AP) component, enabling efficient multi-modal data management, scalable transaction processing, and advanced graph analytics.
\item To demonstrate ArcNeural's query language and API, which provide a unified interface for both TP and AP operations, simplifying the development of complex multi-modal data analysis tasks.
\item To showcase the performance and scalability of ArcNeural through experimental evaluations and real-world use cases, highlighting its potential to support intelligent data applications in various domains.
\end{itemize}
By addressing these objectives, this paper aims to introduce ArcNeural as a powerful and flexible multi-modal database solution for the era of large models, enabling enterprises to effectively leverage their data assets and drive intelligent applications.

\section{Related Work}

Review relevant literature on multi-modal databases and graph databases
Identify gaps in existing research and industrial solutions

\section{ArcNeural: A Multi-Modal Database}
\subsection{System Overview}
ArcNeural is a multi-modal database that employs a storage-compute separated architecture, which aligns with the characteristics of cloud-native environments. This design allows the database to flexibly adapt to users' existing storage solutions while supporting multiple storage modes, enhancing the system's flexibility and scalability. The main challenges faced include handling data diversity and meeting real-time update requirements.

\begin{figure}
    \centering
    \includegraphics[width=1\linewidth]{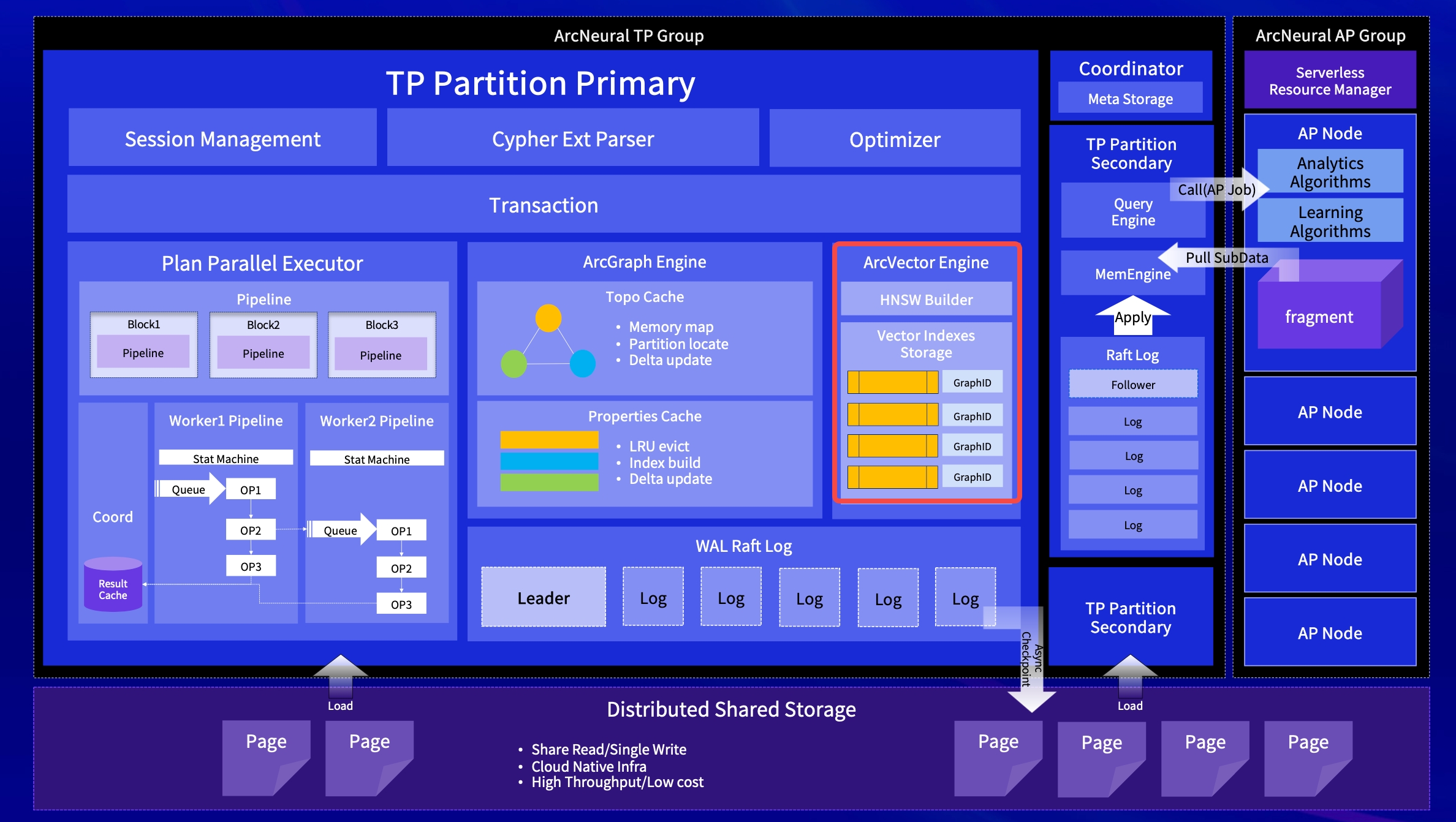}
    \caption{arc-overview}
    \label{fig:arc-overview}
\end{figure}

\subsubsection{Unified Storage Layer}
The storage layer of ArcNeural follows the "log as a database" concept. Storage nodes serve as the persistence layer of a typical database, performing periodic checkpoints by replaying WAL logs. These nodes can be dynamically scaled based on the system's data volume and can be connected to standalone RocksDB, distributed TiKV, or cloud-based object storage systems like OSS.
ArcNeural introduces an innovative design for caching WAL logs at the compute layer, known as the Semi-Stateful concept. The system's WAL logs achieve high availability and consistency among compute nodes using the multi-raft protocol, which implements a distributed consensus algorithm.
\subsubsection{Transaction Processing (TP) Compute Nodes}
Instead of using a single codebase to implement HTAP, ArcNeural considers the distinct usage scenarios of TP and AP in most graph scenarios. At the TP level, ArcNeural implements a complete distributed database system that supports multi-modal engines, currently including graph and vector data models, with plans for further expansion.
\subsubsection{Analytical Processing (AP) Component}
The AP component adopts a serverless architecture, meaning it does not need to run continuously. Instead, it starts analysis nodes in the resource pool on-demand. This design allows data to flow from TP to AP in real-time, and the results can be fed back to TP nodes for subsequent operations like data updates. This mechanism enables the organic integration of TP and AP systems, maximizing their respective advantages.

\subsection{Vector Indexing Integration}

In the era of large language models and generative AI, the efficient storage and retrieval of high-dimensional vector data have become imperative. Vector representations are essential for a variety of AI applications including semantic search, recommendation systems, and retrieval-augmented generation (RAG). Recognizing these needs, ArcNeural has integrated native support for vector data and indexing within its multi-modal database engine.

This integration allows for the seamless storage and querying of vector representations alongside other data types such as graphs and documents. Consequently, developers can exploit the synergies of structured and unstructured data to build more robust and powerful AI applications.

Further enhancing its capabilities, ArcNeural supports the storage of vector representations for diverse types of text data, such as movie content or reviews. This facilitates complex analytical tasks that require nuanced data interpretations. To optimize the retrieval of vector data, ArcNeural employs Hierarchical Navigable Small World (HNSW) indexing, supplemented by Single Instruction, Multiple Data (SIMD) optimizations for rapid distance calculations. Additionally, the system incorporates attribute filtering in vector searches and provides full database functionalities specifically tailored for handling vector data.

\subsubsection{Implementation Details}

Query Language Parsing and Execution Plan:

ArcNeural extends its query language compiler to support the new vector-related syntax.
The compiler introduces new tokens for ARRAY (vector data type) and VECTOR (vector index).
The abstract syntax tree (AST) and execution plan remain consistent with the existing implementation.

Storage Integration:

Vector data is stored in an external vector storage engine or a remote ArcVector cluster.
During the checkpoint process, the leader node of ArcNeural flushes the necessary vector data to the external storage.
The checkpoint process occurs on the ArcNeural server machines and directly interacts with the vector storage engine's API.

Query Execution:

ArcNeural distinguishes between two types of queries: regular queries (without vector indexing) and vector index-based queries (requiring vector distance calculations).
For regular queries, the execution remains similar to other data types, with predefined operations on vector fields (e.g., equality comparison, vector norm calculation).
For vector index-based queries, ArcNeural introduces a new operator called VertexVectorScan that returns the vertex key and the similarity score.
During the optimization phase, the VertexScan operator is replaced with VertexVectorScan when vector indexing is required.
In future iterations, ArcNeural plans to introduce the VertexFilterableIndexScan operator, which supports both vector and scalar filtering, to minimize the impact on regular queries by pushing down filters to the vector storage engine.

\subsubsection{Vector Storage Design}
ArcNeural leverages Qdrant, a vector similarity search engine, to achieve the integration of graph and vector data. The integration is divided into two stages: syntax-level integration and storage-level integration.
In the first stage, ArcNeural focuses on syntax-level integration, utilizing a write-through approach to delegate vector storage capabilities to an external ArcVector service (Qdrant). The basic CRUD operations for vectors have been implemented and tested.
The second stage involves embedding the Filterable HNSW algorithm into ArcNeural's storage system, achieving strong localization of vector index storage. This design focuses on the local implementation of vector storage in the second stage.

Storage Schema Mapping:

ArcNeural maps vertices to collections in the vector storage engine.
Vector fields are mapped to vector information in the collection, while fields with other index types are mapped to payload information.

Local Storage Abstraction:

ArcNeural implements a local storage abstraction layer for vector indexing and storage.
The HNSW index is stored using memory-mapped files (MMAP) for efficient access.
Payload information is stored using RocksDB, a key-value store.

Data Flow:

During the checkpoint process, vector data is written to the local storage.
Bulk insertion (bulk\_upsert) is supported for efficient data ingestion.
Querying supports single-shard reads and merging of data from multiple shards.

Index Construction and Management:

ArcNeural utilizes its job framework to create new jobs for managing segment compaction in the vector storage.
Collection creation and deletion operations are supported through the create\_collection and delete\_collection commands.

Recovery and Fault Tolerance:

Upon process restart, ArcNeural loads the collection cache to restore the vector storage state.
In case of machine failure, ArcNeural relies on snapshots for recovery, following a similar process to graph snapshot build/send/receive/apply.

The integration of vector data and indexing capabilities in ArcNeural enables powerful AI applications that leverage the combined strengths of structured and unstructured data. By providing native support for vector storage and querying, ArcNeural facilitates the development of retrieval-augmented generation and other advanced AI functionalities.

\subsection{Data Model}
ArcNeural's data model seamlessly integrates vectors, documents, and graphs. It leverages the strong expressive power of graph technology and extends the attribute handling capabilities of graphs to include vectors, documents, and JSON. This enables support for complex multi-modal data analysis.
\subsection{Query Language and API}
ArcNeural provides a unified query interface for both TP and AP. It builds upon the Cypher graph query language and incorporates extensions for AP capabilities, such as invoking AP functions using MATCH Call statements. The TP layer performs semantic analysis to determine if AP is required and decomposes operators accordingly.
\section{Major Architectural Innovations}

In this section, we provide four major architectural innovations: Log as a database, mem-engine, mpp, and HTAP.

\subsection{Log as A Database}
ArcNeural adopts the popular "log as a database" concept, where storage nodes serve as the persistence layer, performing periodic checkpoints by replaying WAL logs. The compute layer introduces a novel design for caching WAL logs, known as the Semi-Stateful concept, which utilizes the multi-raft protocol for distributed consistency.

\subsection{MemEngine: Accelerating Topology Processing}

\subsubsection{Key Ideas}

MemEngine, integral to ArcNeural, is designed to elevate query performance through advanced memory storage strategies. It employs graph topology indexes and attribute indexes to ensure rapid and efficient processing across various data types, including structured, semi-structured, and unstructured data.

In traditional graph databases, queries involving vertices, edges, and topology information often depend on storage engine interfaces that access data from disk-based databases or files. This reliance can result in substantial disk I/O overhead, particularly with the random access nature of graph data, leading to numerous physical page misses. To mitigate these inefficiencies, many mainstream database systems, such as Oracle's Keep Cache Buffer Pool and MySQL, implement in-memory caching capabilities. In-memory databases have similarly gained traction due to their ability to significantly enhance data access speeds by minimizing disk I/O operations.

MemEngine is optimized for the unique characteristics of graph data, which can be broadly classified into topology data (of edges) and attribute data (of vertices and edges). Given the complexity of graph topology and the potential for cross-host graph partition is often challenging. We assume that, in most scenarios, the entire graph topology data can be cached in memory.

For attributes of vertices and edges, which resemble the tabular data found in relational databases, MemEngine employs an LRU (Least Recently Used) caching strategy. This method is particularly effective as it targets the dynamic access patterns typical in graph queries, where only specific attributes of a vertex type are frequently accessed. The LRU cache thus prioritizes and retains the most actively accessed data in memory, significantly improving response times.

Additionally, MemEngine supports extensive data manipulation capabilities, ranging from simple updates to complex additions and deletions of vertices and edges. This flexibility ensures swift and adaptable query execution in response to evolving data requirements.

The skewed nature of vertex connections in real-world graphs (e.g., LDBC SF10 graph) poses further challenges, where a minor subset of vertices (super nodes) might have extraordinarily high connectivity. The following is the distribution of such connections.

\begin{figure}
    \centering
    \includegraphics[width=1\linewidth]{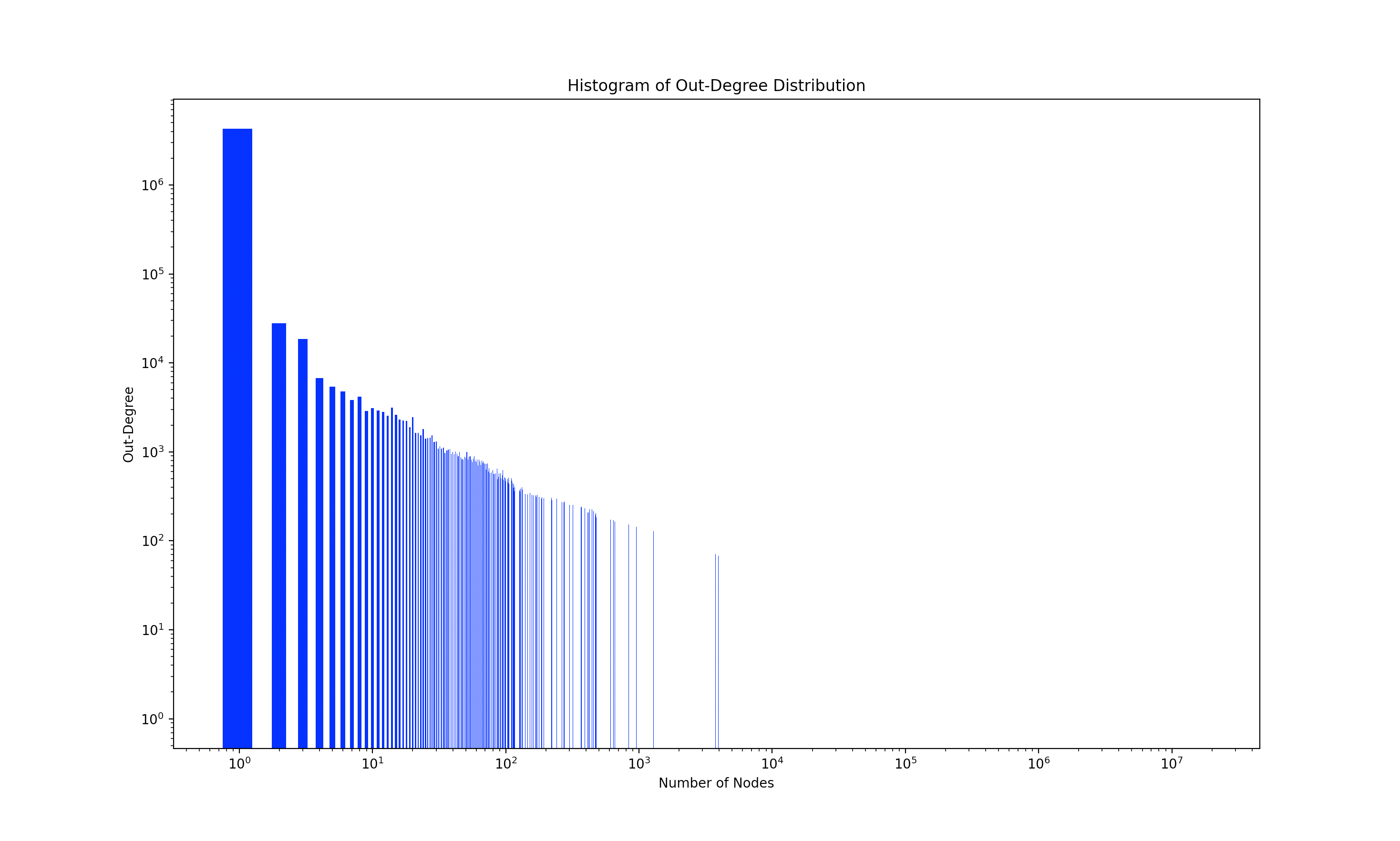}
    \caption{Skewed connections}
    \label{fig:enter-label}
\end{figure}

Moreover, obtaining the degree of a vertex is a common operation in graph traversal and analysis. MemEngine aims to provide high efficiency for this operation, which further emphasizes the need for a specialized edge storage structure.

\subsubsection{Adaptive Edge Collection}

\begin{figure}
    \centering
    \includegraphics[width=1\linewidth]{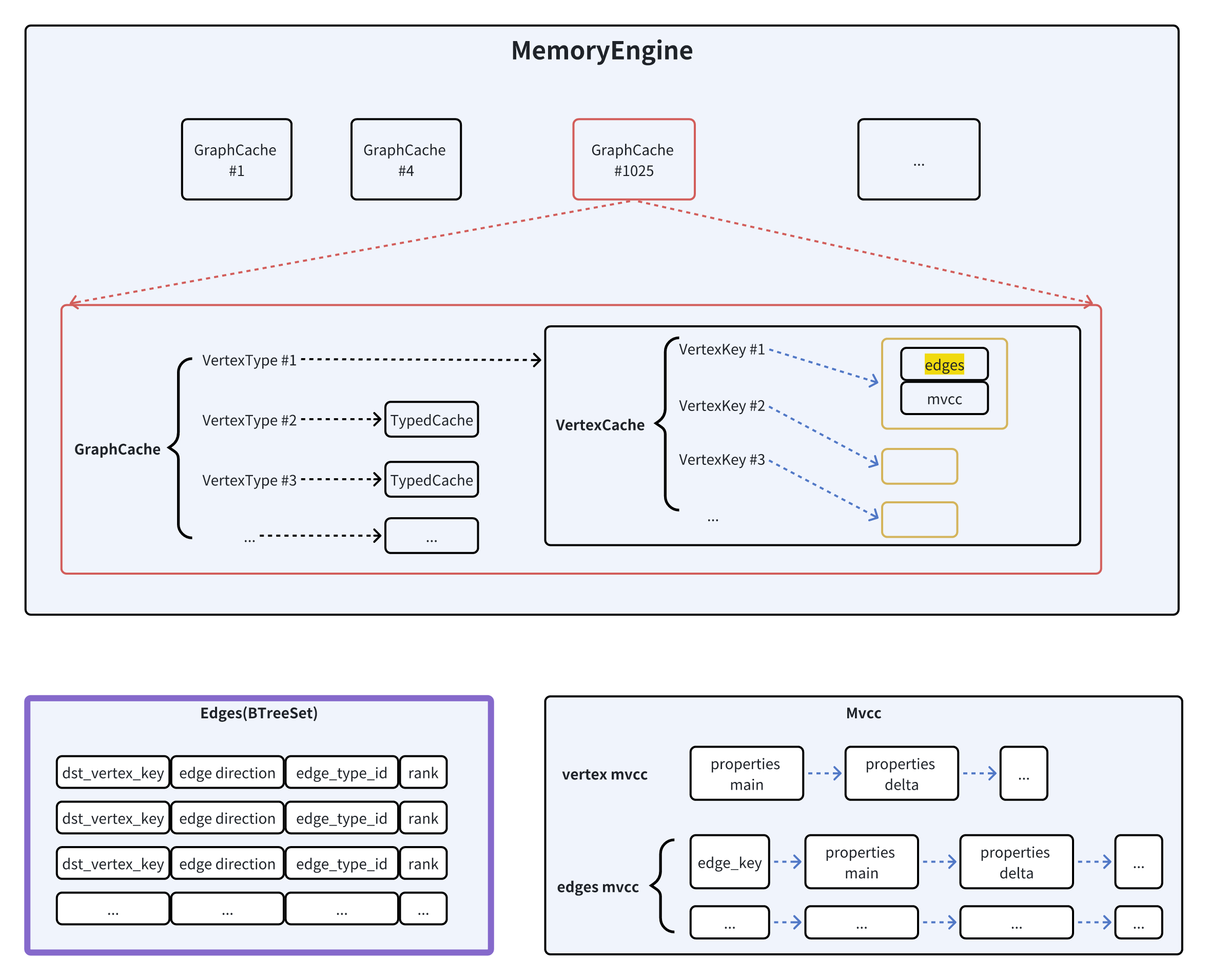}
    \caption{MemoryEngine}
    \label{fig:eMemoryEngine}
\end{figure}

In MemEngine, we introduce a data structure called Adaptive Edge Collection to efficiently store and manage edge information. The Adaptive Edge Collection adapts its internal representation based on the number of edges associated with a vertex, optimizing for both memory usage and query performance.

The Adaptive Edge Collection addresses these challenges by adapting its representation based on the degree of each vertex. For vertices with a low degree, a compact and cache-friendly data structure, such as a vector, is used to store the edges. This allows for efficient memory utilization and fast sequential access to the edges. On the other hand, for vertices with a high degree, the storage automatically switches to a more scalable and hierarchy data structure, such as a B-tree or a hash table, to handle the increased edge density.

By dynamically adapting the storage representation based on the degree of each vertex, MemEngine can efficiently handle graphs with highly skewed degree distributions. This approach optimizes both memory usage and performance, ensuring that the majority of vertices with a low degree are stored compactly while providing efficient access to the edges of super nodes.

Taking inspiration from the implementation of Sets in Redis, we have chosen 128 as the default threshold value. When the number of edges is less than or equal to 128, a \texttt{Vec} is used to store the edges. When the number of edges exceeds 128, a \texttt{BTreeSet} is used instead. This approach balances storage overhead and query performance.

The Adaptive Edge Collection is implemented using a combination of Rust's Vec and BTreeSet data structures, with a configurable threshold to determine when to switch between the two representations. This allows MemEngine to optimize edge storage based on the specific characteristics of the graph and the available memory resources.

Using the Adaptive Edge Collection, MemEngine can efficiently store and retrieve edge information, enabling fast traversal and querying operations on the graph. The adaptive nature of the storage ensures that the system can handle graphs with diverse edge densities, from sparse to highly connected regions, without compromising performance or memory efficiency.

\subsubsection{Performance Metrics}

The memory footprint efficiency of the MemEngine design is demonstrated in Table \ref{tab:memory-footprint}. The table compares the memory usage of different data structures for storing vertices and edges.

\begin{table}[h]
\centering
\adjustboxset{max width=\textwidth}
\begin{adjustbox}{width=1\textwidth}
\begin{tabular}{|l|c|c|c|}
\hline
\textbf{Data Structure} & \textbf{Vertices} & \textbf{\makecell[c]{Vertices +\\Forward Edges}} & \textbf{\makecell[c]{Vertices + Forward\\+ Reverse Edges}} \\
\hline
BTreeSet & 3.8 GB & 10.8 GB & 16.5 GB \\
\hline
Vec & 3.8 GB & 7.3 GB & 10.7 GB \\
\hline
\makecell[l]{Box\textless{Vec}\textgreater + Box\textless{BTreeSet}\textgreater\\(factor = 64)} & 3.9 GB & 8.4 GB & 14.3 GB \\
\hline
\makecell[l]{Box\textless{Vec}\textgreater + Box\textless{BTreeSet}\textgreater\\(factor = 128) (default)} & 3.9 GB & 8.0 GB & 13.6 GB \\
\hline
\makecell[l]{Box\textless{Vec}\textgreater + Box\textless{BTreeSet}\textgreater\\(factor = 256)} & 3.9 GB & 8.0 GB & 13.2 GB \\
\hline
Box\textless{Vec}\textgreater & -- & -- & 11.3 GB \\
\hline
Box\textless{BTreeSet}\textgreater & -- & -- & 17.0 GB \\
\hline
\end{tabular}
\end{adjustbox}
\caption{Memory footprint comparison of different data structures in MemEngine}
\label{tab:memory-footprint}
\end{table}

Table~\ref{tab:query-performance} demonstrates how graph traversal performance is not affected by the implementation of the Adaptive Edge Collection in MemEngine. The data highlights the query times for various traversal operations, comparing the performance of the Originals with the new MemEngine enhancements. This indicates that the Adaptive Edge Collection effectively manages the increased complexity and higher data densities without significantly impacting the overall efficiency of query operations.

\begin{table}[h]
\centering
\adjustboxset{max width=\textwidth}
\begin{adjustbox}{width=1\textwidth}
\begin{tabular}{|p{0.5\textwidth}|c|c|}
\hline
\textbf{Query} & \textbf{Original (ms)} & \textbf{New (ms)} \\
\hline
\makecell[l]{MATCH (m:person)-[e:knows * 2]\\-\textgreater(n:person) RETURN n\\LIMIT 1000;} & \{1942, 1966, 1945, 1963\} & \{1992, 1996, 2106, 2062, 1916\} \\
\hline
\makecell[l]{MATCH (m:person)-[e:knows]\\-\textgreater(n:person) RETURN n\\LIMIT 1000;} & \{422, 413, 404, 477\} & \{435, 420, 416, 418\} \\
\hline
\makecell[l]{MATCH (m:person)\\RETURN m.firstName LIMIT 1000;} & \{127, 130, 133\} & \{128, 135, 142, 127, 132\} \\
\hline
\makecell[l]{MATCH (m:person)\\RETURN m.firstName LIMIT 1000;} & \{121, 143, 126\} & \{105, 106, 96\} \\
\hline
\end{tabular}
\end{adjustbox}
\caption{Query performance comparison between original and new implementations}
\label{tab:query-performance}
\end{table}

TODO

\begin{figure*}[h!]
\centering
\includegraphics[width=\textwidth]{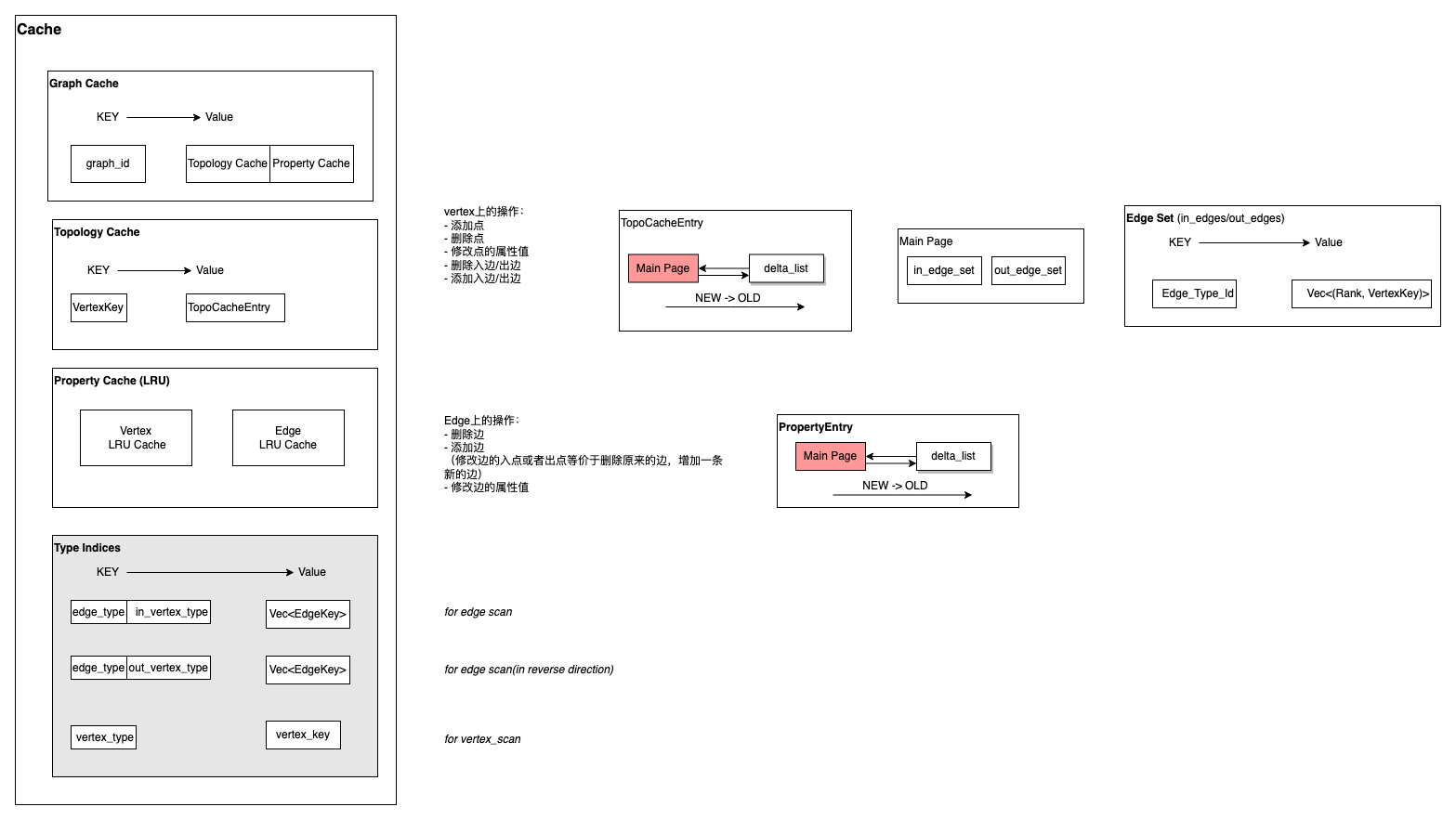}
\caption{Mem-engine-2 TODO}
\label{fig:Mem-engine-2}
\end{figure*}

\begin{figure*}[h!]
    \centering
    \includegraphics[width=\textwidth]{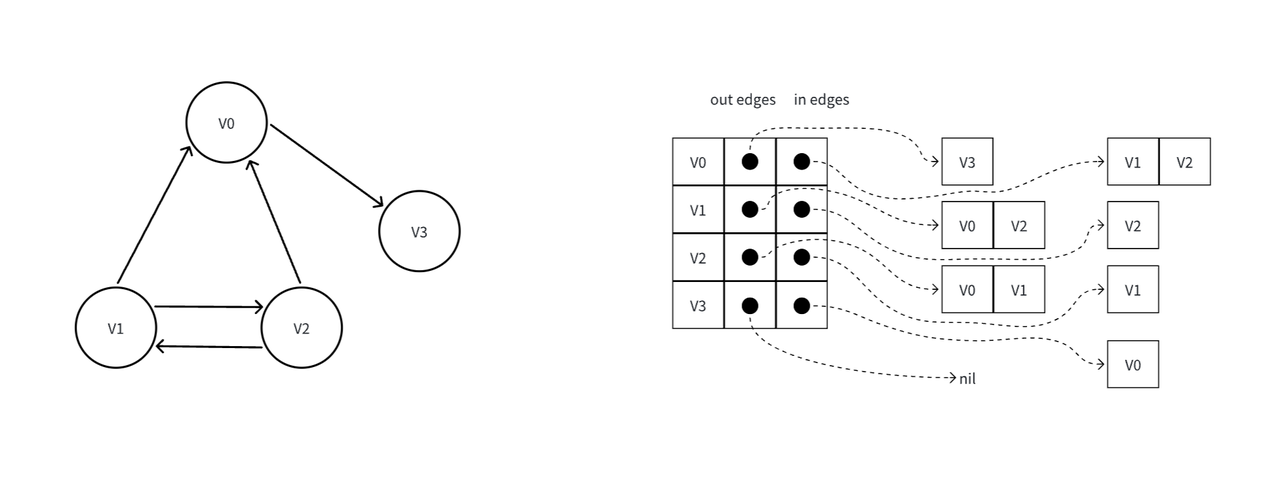}
    \caption{mem-engine-map-structure }
    \label{fig:mem-engine-map-structure}
\end{figure*}

\subsection{Scalable Query Execution}
The compute query layer adopts an MPP distributed architecture, with each compute node using a push-based vectorized execution engine. The execution plan is split based on operator attributes, generating a DAG graph. The plan is then distributed to the involved partition leader nodes for distributed parallel execution.
\subsection{Unified HTAP}
ArcNeural achieves the fusion of TP and AP by providing a unified query interface. The TP layer, based on the Cypher query language, is extended to support AP capabilities. When a request is received, the TP layer performs semantic analysis to determine if AP is needed, decomposes operators, prepares data, and invokes AP accordingly.
\section{Benchmarking}

Experimental Evaluation
\subsubsection{Experimental Setup}
- Describe the hardware and software configuration used for experiments
\subsubsection{Datasets and TP,AP,vector Workloads}
- Introduce the datasets and workloads used for evaluation
\subsubsection{Performance Results}
- Present and analyze the performance results
- Compare ArcNeural's performance with other state-of-the-art systems

\section{Industrial Applications}

Discuss real-world industrial applications where ArcNeural has been deployed
Highlight the benefits and impact of using ArcNeural in these scenarios

\section{Conclusions \& Future Work}

Summarize the main contributions and findings of the paper
Discuss limitations and potential future research directions

\section*{Acknowledgments}

\bibliography{sn-bibliography}


\section{Appendices}

\end{document}